\documentclass[conference,10pt]{IEEEtran}
\usepackage{diagbox} 
\usepackage{graphicx}
\usepackage{epstopdf}
\usepackage{psfrag}
\usepackage{subfigure}
\usepackage{url}
\usepackage{stfloats}
\usepackage{amsfonts,amssymb,amsmath,bm,paralist,theorem,cite,ifthen,color,nccmath}
\usepackage{caption}
\usepackage{calc}
\usepackage{enumerate}
\usepackage{multirow}
\usepackage{makecell}
\usepackage[ruled]{algorithm2e}

\usepackage{array}
\usepackage{float}
\usepackage{soul}
\usepackage[colorlinks,
            linkcolor=blue,
            anchorcolor=red,
            citecolor=red]{hyperref}

\graphicspath{{Figures/}}

\newcommand{\T}{{\scriptscriptstyle\mathsf{T}}}
\renewcommand{\H}{{\scriptscriptstyle\mathsf{H}}}

\newcommand\Ccl{\ensuremath{\mathcal{C}}}
\newcommand\Dcl{\ensuremath{\mathcal{D}}}

\newcommand\Ncl{\ensuremath{\mathcal{N}}}

\newcommand\Scl{\ensuremath{\mathcal{S}}}
\newcommand\Icl{\ensuremath{\mathcal{I}}}

\newcommand\Vcl{\ensuremath{\mathcal{V}}}

\newcommand\Zcl{\ensuremath{\mathcal{Z}}}

\newcommand\Lcl{\ensuremath{\mathcal{L}}}

\newcommand\Cs{\ensuremath{{\mathbb{C}}}}
\newcommand\Es{\ensuremath{{\mathbb{E}}}}
\newcommand\Rs{\ensuremath{{\mathbb{R}}}}

\newcommand\Pbb{\ensuremath{{\mathbb{P}}}}

\newcommand\Pb{\ensuremath{ \mathbf{P} }}

\newcommand\Sb{\ensuremath{ \mathbf{S} }}

\newcommand\Zb{\ensuremath{ \mathbf{Z} }}

\newcommand\bb{\ensuremath{ \mathbf{b} }}

\newcommand\fb{\ensuremath{ \mathbf{f} }}
\newcommand\hb{\ensuremath{ \mathbf{h} }}

\newcommand\pb{\ensuremath{ \mathbf{p} }}

\newcommand\vb{\ensuremath{ \mathbf{v} }}

\newcommand\Nr{\ensuremath{ N_{\rm r} }}
\newcommand\Nc{\ensuremath{ N_{\rm c} }}
\newcommand\Ns{\ensuremath{ N_{\rm s} }}

\usepackage[labelsep=period,font=footnotesize]{caption}

\usepackage{etoolbox}
\newcommand{\zerodisplayskips}{%
  \setlength{\abovedisplayskip}{3pt}%
  \setlength{\belowdisplayskip}{3pt}%
  \setlength{\abovedisplayshortskip}{3pt}%
  \setlength{\belowdisplayshortskip}{3pt}}
\appto{\normalsize}{\zerodisplayskips}
\appto{\small}{\zerodisplayskips}
\appto{\footnotesize}{\zerodisplayskips}

\usepackage{titlesec}
\titlespacing{\section}{-0.64 cm}{2pt}{2pt}
\titlespacing{\subsection}{0 cm}{2pt}{2pt}
\usepackage[normalem]{ulem}

\usepackage[all=normal,paragraphs=tight,floats=normal,mathspacing=normal,wordspacing=tight,charwidths=tight,mathdisplays=normal,leading=normal]{savetrees}
\IEEEoverridecommandlockouts
\usepackage{xcolor}

\begin{document}

\title{\vspace{-8mm}Knowledge Distillation for Lightweight Multimodal Sensing-Aided mmWave Beam Tracking\vspace{-5mm}}
\author{
\IEEEauthorblockN{Mengyuan Ma$^*$, Isuri Welgamage$^*$, Ahmed Alkhateeb$^\dagger$, A.~Lee~Swindlehurst$^\S$,\\ Markku Juntti$^*$, and Nhan Thanh Nguyen$^*$}
\IEEEauthorblockA{$^*$Centre for Wireless Communications (CWC), University of Oulu, Finland \\
$^\dagger$School of Electrical, Computer, and Energy Engineering, Arizona State University, Tempe, USA\\
$^\S$School of Electrical Engineering \& Computer Science, University of California, Irvine, USA\\
Email: \{mengyuan.ma, isuri.welgamage, nhan.nguyen, markku.juntti\}@oulu.fi;alkhateeb@asu.edu;swindle@uci.edu
}\vspace{-10mm}}

\maketitle

\begin{abstract}
Beam training and prediction in real-world millimeter-wave (mmWave) communications systems are challenging due to rapidly time-varying channels and strong interference from surrounding objects. In this context, widely available sensors, such as cameras and radars, can capture rich environmental information, enabling efficient beam management. This paper proposes a knowledge-distillation (KD)-enabled learning framework for developing lightweight and low-complexity models for beam prediction and tracking using real-world camera and radar data from the DeepSense 6G dataset. Specifically, a powerful teacher network based on convolutional neural networks (CNNs) and gated recurrent units (GRUs) is first designed to predict current and future beams from historical sensor observations. Then, a compact student model is constructed and trained via KD to transfer the predictive capability of the teacher model to a lightweight architecture. Simulation results demonstrate that jointly leveraging radar and image modalities significantly outperforms single-modality approaches. Moreover, the proposed student model achieves over 96\% Top-5 beam prediction accuracy while reducing computational complexity by more than $4$ times and the number of parameters by over $27$ times compared with the teacher model.
\end{abstract}


\IEEEpeerreviewmaketitle

\section{Introduction}
Millimeter-wave (mmWave) communications systems enable extremely high data rates for emerging applications such as vehicular networks, unmanned aerial vehicles, and augmented/virtual reality \cite{jiang2021road}. However, severe path loss at these frequencies necessitates large antenna arrays to form narrow directional beams toward user equipment. Conventional beam training typically relies on exhaustive searches over large beam codebooks, resulting in significant overhead in highly dynamic environments \cite{imran2024environment}. A promising direction is to exploit low-cost sensors, such as radar and cameras, to capture environmental information and enable sensing-assisted beam prediction and alignment.

Existing studies have demonstrated the feasibility of sensing-assisted beam prediction using position \cite{morais2023position}, radar \cite{Demirhan2022Radar}, vision \cite{imran2024environment}, and LiDAR \cite{marasinghe2022lidar} data. However, relying on a single sensing modality may lead to degraded performance under varying environmental conditions. For instance, vision-based sensing is sensitive to lighting and weather conditions. Therefore, combining multiple sensing modalities is attractive because it can provide complementary environmental information and improve system robustness. Several works have explored multimodal sensing for beam prediction and alignment. Charan~{\it et al.} \cite{charan2022vision} combined vision and GPS data for beam prediction and showed that multimodal sensing outperforms single-modality approaches. More recent studies have investigated advanced multimodal feature fusion and sensing-assisted beam prediction frameworks \cite{cui2024sensing,zhu2025advancing,park2025resource}.

Despite these advances, most existing works focus on predicting the beam for only the current time slot using sensory observations, while overlooking model complexity. Such approaches require frequent inference at every time step, leading to high computational complexity and substantial sensing and processing overhead. Long-term beam prediction can mitigate these issues by predicting the beams for not only the current but also future time slots. However, existing studies on long-term beam prediction mainly rely on single-modality sensing, such as LiDAR \cite{Jiang2024LiDAR}, radar \cite{Luo2023millimeter}, and vision \cite{ma2025attention,ma2025knowledge}, and rarely consider model complexity. Using multimodal sensing for efficient beam tracking remains largely unexplored.

In this work, we investigate long-term beam tracking using both vision and radar sensing data and propose an efficient knowledge distillation (KD)-aided teacher–student learning framework for developing lightweight models with strong prediction performance. Specifically, we first design a high-capacity teacher model that integrates convolutional neural networks (CNNs), gated recurrent units (GRUs), and a multi-head attention (MHA) mechanism to capture spatial and temporal features for long-term beam tracking. Then, we construct a lightweight student model using depthwise separable convolution \cite{howard2017mobilenets} and train it with guidance from the teacher model via KD. To the best of the authors’ knowledge, this is the first work that employs multimodal sensing for long-term beam tracking while reducing model complexity. Experiments on the DeepSense 6G dataset demonstrate that the teacher model achieves over $96\%$ Top-5 beam prediction accuracy across the current and the next three time slots, outperforming the state-of-the-art radar-only and vision-only baselines. Moreover, the student model approaches the teacher’s performance while reducing the model size by $27\times$ and computational complexity by $4\times$.

\section{System Model and Problem Formulation}\label{sec:system model}
 \begin{figure*}[t]
\vspace{-9mm}
	\small
	\centering	
	\includegraphics[width=1\textwidth]{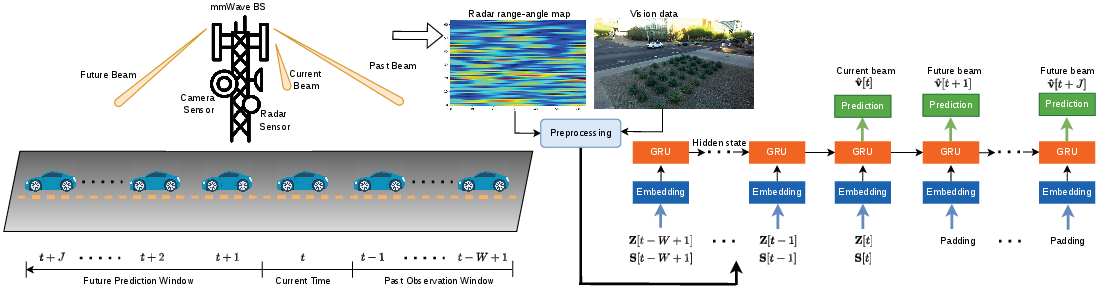}
	\vspace{-4mm}
	\caption{System illustration.}
	\label{fig:system model}
	\vspace{-5mm}
\end{figure*}
We consider a downlink mmWave communications system in which a base station (BS) equipped with a uniform linear array serves a single-antenna mobile user equipment (UE), as illustrated in Fig.~\ref{fig:system model}. In addition to the communications array, the BS is equipped with an RGB camera and a frequency-modulated continuous wave (FMCW) radar sensor to capture real-time environmental dynamics.

\subsection{System Model}\label{sec:system model_A}

{\bf Communication model:} 
At time slot $t$, the BS transmits a symbol $s[t]\in \Cs$ with $\Es\!\left[|s[t]|^2\right]=1$ to the UE. We assume a block fading channel between time slots. Let $\vb[t]$ denote the beamforming vector at time slot $t$. The received signal $y[t]$ at the UE can be expressed as
\begin{align}\label{eq:signal model}
y[t] = \hb[t]^\H \vb[t] s[t] + n[t],
\end{align}
where $\hb[t]$ denotes the channel between the BS and the UE at time slot $t$, and $n[t]\sim\Ccl\Ncl(0,\sigma_{\rm n}^2)$ represents additive white Gaussian noise (AWGN) with power $\sigma_{\rm n}^2$. The signal-to-noise ratio (SNR) at time slot $t$ is given by $\text{SNR}[t] = \frac{|\hb[t]^\H \vb[t]|^2}{\sigma_{\rm n}^2}$.

{\bf Radar model:} 
The FMCW radar transmits periodic chirp signals and receives the reflected echoes from surrounding objects. After quadrature mixing and low-pass filtering, the received radar signal is converted to an intermediate-frequency (IF) signal and sampled by an analog-to-digital converter (ADC). For each sensing frame, the radar transmits $\Nc$ chirps and collects $\Ns$ fast-time samples per chirp at each of the $\Nr$ receive antennas. The radar observations at time $t$ can therefore be represented as a three-dimensional tensor $\mathbf{S}[t] \in \mathbb{C}^{\Nr \times \Ns \times \Nc}$,
where the three dimensions correspond to the antenna, fast-time (range), and slow-time (chirp) domains, respectively. The radar observations capture the spatial and temporal dynamics of surrounding objects, which can be leveraged together with visual sensing data to assist beam prediction and tracking for the UE.

\subsection{Problem Formulation}
At time slot $t$, our goal is to determine the transmit  beamforming vectors at the BS for the current and $J$ future time slots $\{t, t+1, \ldots, t+J\}$. Let $\Vcl = \{\vb_1,\ldots, \vb_{C}\}$ and $\Icl_{\Vcl}=\{1,\ldots,C\}$ denote the beamforming codebook and its associated index set with $C\triangleq|\Vcl|$. In the considered beam tracking problem, we aim to find $\vb[\tau] \in \Vcl, \forall \tau$, to maximize the spectral efficiency $R_J=\sum_{\tau=t}^{t+J}\log\left(1+ \text{ SNR}[\tau]\right)$ over the $J+1$ time slots. 
For low SNR, we can formulate the beamforming problem as \cite{Jiang2024LiDAR,Luo2023millimeter,ma2025attention,ma2025knowledge}
\begin{equation}\label{pb:P2}
    \underset{\vb[\tau] \in \Vcl, \forall \tau}{\rm maximize} \quad  \sum_{\tau=t}^{t+J}|\hb[\tau]^\H \vb[\tau]\big|^2.
\end{equation}
Let $\bb^{\star}[t]=\big[b^{\star}[t],\ldots,b^{\star}[t+J]\big]^\T$ be the vector of beam indices corresponding to the optimal solution of \eqref{pb:P2}, i.e., 
\begin{equation}\label{pb:P3}
      \bb^{\star}[t] = \underset{b[\tau]\in \Icl_{\Vcl},\forall \tau}{\arg\max} \quad \sum_{\tau=t}^{t+J}|\hb[\tau]^\H \vb_{b[\tau]}\big|^2.
\end{equation}
In principle, the solution to \eqref{pb:P3} could be obtained by decoupling it into $J+1$ subproblems where each is solved via an exhaustive search over the $C$ candidate beams. However, the complexity of such a method scales as $J\cdot C$, which incurs high latency, especially with the large codebooks used in massive MIMO. Moreover, this approach requires perfect channel state information (CSI) at not only the current time slot, but also the $J$  future time slots, which is generally unavailable in practice.

We therefore consider \emph{CSI-free} beam tracking: instead of estimating $\hb[\tau]$, we leverage sensed visual and radar observations, denoted by $\Zb[t]$ and $\Sb[t]$, to predict the beam-index sequence $\bb[t]$ such that the solution remains effective with respect to the CSI-based objective in \eqref{pb:P3}. Unlike prior works that solve \eqref{pb:P3} via per-slot decoupling \cite{Demirhan2022Radar,morais2023position,imran2024environment,marasinghe2022lidar,charan2022vision,cui2024sensing,zhu2025advancing,park2025resource}, we develop a learning framework that \emph{directly} predicts the entire horizon $\bb[t]$ for long-term beam tracking.

\section{Multimodal Sensing for Beam Tracking}
\subsection{Learning Task}
Let $\Zb[t]\in\Rs^{3\times d_{\rm H}\times d_{\rm W}}$ denote the RGB image obtained at time slot $t$, where the dimension $3$ corresponds to the number of RGB color channels, and $d_{\rm H}$ and $d_{\rm W}$ respectively represent the image height and width in pixels. Define $\Zcl[t]=\{\Zb[t-W],\Zb[t-W+1], \ldots,\Zb[t]]\}$ and $\Scl[t]=\{\Sb[t-W],\Sb[t-W+1], \ldots,\Sb[t]]\}$ as the sequence of raw sensory images and radar data, respectively, from $W$ past time slots to the current time $t$. We aim to develop an efficient machine learning (ML) model to solve \eqref{pb:P2}, i.e., to predict the optimal beams (equivalently the optimal beam indices in $ \Icl_{\Vcl}$) for the current time slot $t$ and the $J$ future time slots $t+1,\ldots, t+J$. Let $f(g(\Zcl[t]), r(\Scl[t]);\Theta)$ denote the ML model with learnable parameters $\Theta$, where $g(\cdot)$ and $r(\cdot)$ respectively represent the data preprocessing operations for raw images and radar data. The ML model outputs the probabilities of all possible beams at the $J+1$ current and future time slots. Let $p_c[t + j]$ denote the probability of selecting the $c$-th beam in the codebook at time $t + j$, and define $\pb[t + j] = [p_1[t + j], \ldots, p_{C}[t + j]]^\T \in \Rs^{C}, j=0,\ldots,J$. 
The predicted beam index is obtained as
\begin{equation}\label{eq:predicted beam index}
    \hat{b}[\tau]=\arg\max_{c\in\Icl_{\Vcl}} \; p_c[\tau] ,\ \tau=t,\ldots,t+J.
\end{equation}
The desired ML model for vision and radar-aided beam tracking is written as
\begin{equation}\label{pb:ML task}
    \Theta^{\star}=\arg\max_{\Theta} \quad \sum_{\tau=t}^{t+J} \Pbb\{\hat{b}[\tau]=b^{\star}[\tau]\},
\end{equation}
where $\Pbb\{\cdot\}$ denotes an event probability.  We note that $J$ and $W$ are empirically determined hyperparameters.

\subsection{Beam Tracking ML Model Structure}\label{sec:model structures}

As seen in \eqref{pb:ML task}, the considered learning task is to predict a sequence of beam selections over the current and future time steps based on a time series of past vision and radar data. 
Instead of employing computationally heavy Transformer-based architectures as in \cite{cui2024sensing,zhu2025advancing} to solve such a sequence-to-sequence (Seq2Seq) learning problem, we develop an efficient framework based on GRU and attention mechanisms.

 \begin{figure}[t]
\vspace{-8mm}
	\small
	\centering	
	\includegraphics[width=0.5\textwidth]{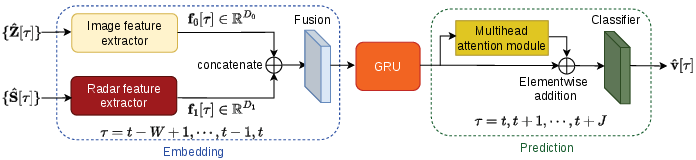}
	\vspace{-5mm}
	\caption{Beam tracking model structure.}
	\label{fig:Model structure}
	\vspace{-2mm}
\end{figure}

Fig.~\ref{fig:Model structure} illustrates the considered GRU-based Seq2Seq architecture, which consists of three functional blocks: embedding, temporal modeling, and prediction. The embedding block extracts low-dimensional semantic features from high-dimensional image and radar inputs. Specifically, let $\{\hat{\Zb}[\tau]\}$ and $\{\hat{\Sb}[\tau]\}$ denote the preprocessed image and radar sequences, respectively. These inputs are fed into their corresponding feature extractors to obtain feature vectors $\{\fb_0[\tau]\in\mathbb{R}^{D_0}\}$ and $\{\fb_1[\tau]\in\mathbb{R}^{D_1}\}$. The extracted features are concatenated as
\begin{equation}
\mathbf{f}[\tau]=
\begin{bmatrix}
\mathbf{f}_0[\tau]\\
\mathbf{f}_1[\tau]
\end{bmatrix}
\in\mathbb{R}^{D_0+D_1},\quad \tau=t-W+1,\ldots,t,
\end{equation}
and then projected into a latent space of dimension $D$ through a multi-layer perceptron (MLP) fusion layer. The resulting feature sequence is processed by a GRU to capture temporal dependencies. Finally, the prediction block generates the predicted beam vectors $\hat{\vb}[t+j]$, $j=0,\ldots,J$, using a multi-head attention (MHA) module \cite{vaswani2017attention} followed by an MLP classifier with $C$ outputs. By applying an MHA after the GRU, the model can exploit global dependencies across the sequence, complementing the GRU's temporal modeling capability and yielding richer feature representations.


\textbf{Teacher Model Structure:} To achieve strong predictive performance, we first develop a high-capacity teacher model based on the above architecture. Instead of using large pretrained backbones such as ResNet \cite{cui2024sensing,park2025resource}, we design task-specific CNN embedding networks for image and radar feature extraction. The teacher employs standard convolutional layers, a two-layer GRU, and an MHA-based prediction module, which together provide strong spatial feature extraction and temporal modeling for accurate multimodal beam prediction. This performance, however, comes at the cost of high complexity, i.e., about $2.931$M parameters and $179.248$M FLOPs.

\textbf{Student Model Structure:} To reduce the computational complexity of the learning task, we propose a lightweight student model as a compact version of the teacher and train it via KD. Specifically, the student replaces standard convolutions with depthwise separable convolutions (DS-Conv) \cite{howard2017mobilenets} and uses a single-layer GRU. As a result, the student requires only $0.106$M parameters and $42.723$M FLOPs, corresponding to over $27$ times fewer parameters and more than $4$ times lower complexity than the teacher. However, these simplifications also weaken the student's representational and temporal modeling capability, making accurate long-term beam prediction more challenging. To address this challenge, we leverage KD to incorporate the teacher's soft supervision into student training. Such teacher knowledge facilitates student learning and helps mitigate the performance loss caused by the reduced model capacity. The details of training the student model with KD are elaborated on below.

\subsection{Training Student Model with KD}\label{sec:KD aided learning}
Let $\Lcl_{\rm task}$ denote the task loss computed from the dataset and $\Lcl_{\rm distill}$ the distillation loss measuring the discrepancy between teacher and student outputs. The student model is trained by minimizing the overall loss
\begin{align}
\label{eq:overall loss}
\Lcl_{\rm s}=(1-\beta)\Lcl_{\rm task}+\beta\Lcl_{\rm distill},
\end{align}
where $\beta\in[0,1]$ balances supervision from ground-truth labels and teacher outputs. 

{\bf Task Loss:} Let $\Dcl=\{\{\Zcl[t],\Scl[t],\bb^\star[t]\},t=0,\ldots,T\}$ denote the dataset, where $\Zcl[t]$ and $\Scl[t]$ are the input observations and $\bb^\star[t]$ is the optimal beam label obtained from \eqref{eq:signal model}. Because beam labels are unevenly distributed across the $C$ candidate beams, we employ the focal loss \cite{lin2017focal} to address class imbalance. The focal loss for the sample $\{\Zb[\tau],\Sb[\tau]\}$ is given by
\begin{equation}\label{eq:Focal loss}
l_{\rm Focal}[\tau]
=-\alpha(1-p_{b^\star}[\tau])^\gamma
\log\big(p_{b^\star}[\tau]\big),
\end{equation}
where $p_{b^{\star}}[\tau]$ denotes the predicted probability of selecting the ground-truth beam index $b^{\star}[\tau]$ at time slot $\tau$. The hyperparameter $\alpha$ is the weighting factor addressing class imbalance, and $\gamma$ is the focusing parameter that demphasizes easy examples. The overall task loss 
is expressed as
\begin{equation}\label{eq:task loss}
    \Lcl_{\rm task}[t]=\sum_{\tau=t}^{t+J} l_{\rm Focal}[\tau]. 
\end{equation}

{\bf Distillation Loss:} We adopt the Kullback-Leibler (KL) divergence as the distillation loss $\Lcl_{\rm distill}$, which measures the similarity between the output distributions of the teacher and student models, respectively denoted as $\Tilde{P}_{\rm teacher}^{(\tau)}$ and $\Tilde{P}_{\rm student}^{(\tau)}$, given the input sample $\{\Zb[\tau],\Sb[\tau]\}$. The KL divergence is computed as
\begin{equation}
    D_{\rm KL}\left( \Tilde{P}_{\rm teacher}^{(\tau)} \| \Tilde{P}_{\rm student}^{(\tau)} \right) = \sum_{c=1}^C \Tilde{P}_{\rm teacher}^{(\tau,c)} \log \left(\frac{\Tilde{P}_{\rm teacher}^{(\tau,c)}}{\Tilde{P}_{\rm student}^{(\tau,c)}} \right),
\end{equation}
where $\Tilde{P}_{\rm teacher}^{(\tau,c)}$ and $\Tilde{P}_{\rm student}^{(\tau,c)}$ are the predicted probabilities of the $c$-th candidate beam from the teacher and student models, respectively. The distillation loss for the input sequence $\{\Zcl[\tau],\Scl[\tau]\}$ is given by
\begin{equation}\label{eq: distillation loss}
    \Lcl_{\rm distill }[t]= \sum_{\tau=t}^{t+J}  D_{\rm KL}\left( \Tilde{P}_{\rm teacher}^{(\tau)} \| \Tilde{P}_{\rm student}^{(\tau)} \right) \cdot \Gamma^2,
\end{equation}
where the multiplication by $\Gamma^2$ arises since the gradients produced by the softmax function are scaled by $1/\Gamma$. 

\begin{algorithm}[t]
\small
\caption{KD-Aided Training for Problem~\eqref{pb:ML task}.}
\label{alg1}
\LinesNumbered
\KwIn{Training set $\Dcl_{\rm tr}$, validation set $\Dcl_{\rm evl}$, \\ \hspace*{2.8em} and pretrained teacher model $f_{\rm T}(\cdot;\Theta_{\rm T})$}
\KwOut{Optimized student parameters $\Theta_{\rm S}^{\star}$}

\parbox[t]{0.92\linewidth}{
Initialize student parameters $\Theta_{\rm S}$, best model $\Theta_{\rm S}^{\star}$,
and best validation loss $\Lcl_{\rm evl}^{\star}$.
}

\For{$e=1,\ldots,E$}{
    Divide $\Dcl_{\rm tr}$ into mini-batches $\{\Dcl_{\rm tr}^{(n)}\}_{n=1}^{N_{\rm b}}$.
    
    \For{each mini-batch $\Dcl_{\rm tr}^{(n)}$}{
        \parbox[t]{0.92\linewidth}{
        Pre-process the input sequences and obtain\\
        $\Pb_{\rm S}=f_{\rm S}(\cdot;\Theta_{\rm S})$.
        }
        
        \parbox[t]{0.92\linewidth}{
        Obtain teacher predictions $\Pb_{\rm T}=f_{\rm T}(\cdot;\Theta_{\rm T})$.
        }
        
        \parbox[t]{0.92\linewidth}{
        Compute $\Lcl_{\rm task}$, $\Lcl_{\rm distill}$, and $\Lcl_{\rm s}$ 
       using \eqref{eq:task loss}, \eqref{eq: distillation loss}, and \eqref{eq:overall loss}.
        }
        
        Update $\Theta_{\rm S}$ using backpropagation and the optimizer.
    }
    
    \parbox[t]{0.92\linewidth}{
    Evaluate student model on $\Dcl_{\rm evl}$ and obtain $\Lcl_{\rm evl}^{(e)}$.
    }
    
    \If{$\Lcl_{\rm evl}^{(e)} < \Lcl_{\rm evl}^{\star}$}{
        \parbox[t]{0.92\linewidth}{
        Update $\Theta_{\rm S}^{\star}=\Theta_{\rm S}$ and $\Lcl_{\rm evl}^{\star}=\Lcl_{\rm evl}^{(e)}$
        }
    }
}
\Return{$\Theta_{\rm S}^{\star}$}
\end{algorithm}

Algorithm~\ref{alg1} summarizes the KD-aided training procedure. Here, $f_{\rm T}(\cdot;\Theta_{\rm T})$ denotes the pretrained teacher model, while $\Dcl_{\rm tr}$ and $\Dcl_{\rm evl}$ represent the training and validation datasets obtained without overlap from the overall dataset $\Dcl$. The student model parameters $\Theta_{\rm S}$ are first randomly initialized. During each epoch, the training dataset $\Dcl_{\rm tr}$ is divided into mini-batches, and the student model is updated in a batch-wise manner. For each mini-batch, the input sensory sequences are first preprocessed and fed into the student network to obtain the predicted beam probabilities. In parallel, the teacher model produces the corresponding soft predictions. Based on these outputs, the task loss and distillation loss are computed according to \eqref{eq:task loss} and \eqref{eq: distillation loss}, respectively. The overall loss in \eqref{eq:overall loss} is then minimized via backpropagation to update the student model parameters. After each epoch, the student model is evaluated on the validation dataset $\Dcl_{\rm evl}$. The optimal model parameters $\Theta^{\star}_{\rm S}$ are updated if a lower validation loss is found, as shown in steps 11--13. The training is terminated when a predefined maximum number of epochs is reached or the best validation loss $L_{\rm evl}^{\star}$ stops decreasing over a predefined number of consecutive epochs.

\section{Numerical Results and Discussion}\label{sec:simulation}

In this section, we evaluate the performance of the proposed ML model for beam tracking. Experiments are based on Scenario 9 of the DeepSense 6G dataset \cite{alkhateeb2023deepsense}, which provides sensory data and optimal beams for real-world mmWave communications. For any time step $t$, the maximum number of past images is set to $W=8$ in each sequence sample $\{\Zcl[t],\Scl[t]\}$, while the number of future time steps for beam prediction is set to $J=3$. The overall dataset contains a total of $T=4060$ samples $\{\Zcl[t], \Scl[t], \bb^{\star}[t]\}$ with $80\%$ training samples and $20\%$ validation samples. 

{\bf Training Setup:} During training, the initial learning rate is $10^{-4}$ using a cyclic cosine annealing scheduler. We set $\gamma=2$ for the loss function. To improve learning efficiency, the image sequences are first preprocessed by suppressing background interference through adjacent-frame subtraction and highlighting the moving UE using binary motion masks \cite{ma2025knowledge}. For radar sensing, fast Fourier transforms (FFTs) are applied to the raw radar data to generate range-angle and range-Doppler maps, which serve as inputs to the radar feature extractor. Further details on model architectures and experimental setup are available in the released source code.\footnote{The source code is available online at \url{https://github.com/WillysMa/KD-for-sensing.git}.}


{\bf Evaluation Metrics:} We use the Top-$k$ accuracy and distance-based accuracy (DBA) \cite{ma2025knowledge} for performance evaluation. The Top-$k$ accuracy measures whether the ground-truth label is among the model's Top-$k$ predicted labels. The DBA metric computes the distance between the predicted and ground-truth beams and assigns scores based on how far apart they are. The DBA score is computed using Top-$3$ accuracy. Note that both the Top-$k$ and DBA scores target one time slot. To reflect the overall performance across all $J+1$ time slots, we also compute the average Top-$k$ (ATop-$k$) and average DBA (ADBA).

\begin{table}[t]
\centering
\setlength{\tabcolsep}{3pt}
\caption{Overall generalization performance (\%) averaged across time slots.}
\label{tb:overall_generalization}
\begin{tabular}{r|c|c|c|c|c}
\hline
Metric & Radar \cite{Luo2023millimeter} & Image \cite{ma2025attention} & Teacher & \makecell{Student \\ (No KD)} & \makecell{Student \\ (With KD)} \\
\hline
ATop-$3$ & $74.51$ & $83.25$ & $\bf 86.96$ & $80.49$ & $\bf 84.57$ \\
\hline
ATop-$5$ & $90.71$ & $95.03$ & $\bf 96.84$ & $93.57$ & $\bf 95.64$ \\
\hline
ADBA & $91.87$ & $95.46$ & $\bf 96.55$ & $93.97$ & $\bf 95.66$ \\
\hline
\end{tabular}
\end{table}

 \begin{figure}[t]
	\small
	\centering	
	\includegraphics[width=0.35\textwidth]{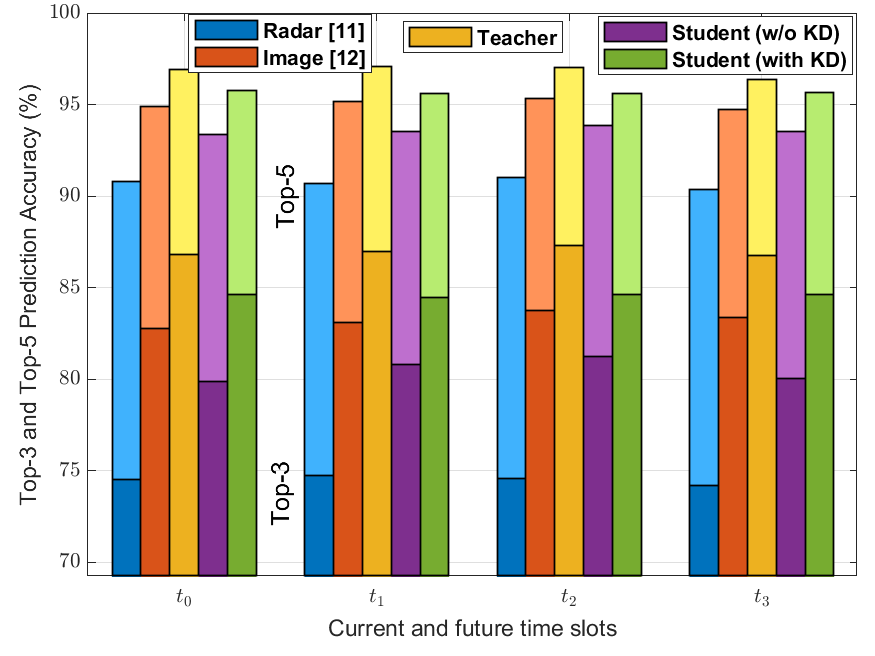}
	\vspace{-2mm}
	\caption{Top-3 and Top-5 prediction accuracy.}
	\label{fig:Top3n5}
\end{figure}

\begin{figure}[t]
\vspace{-5mm}
	\small
	\centering	
	\includegraphics[width=0.35\textwidth]{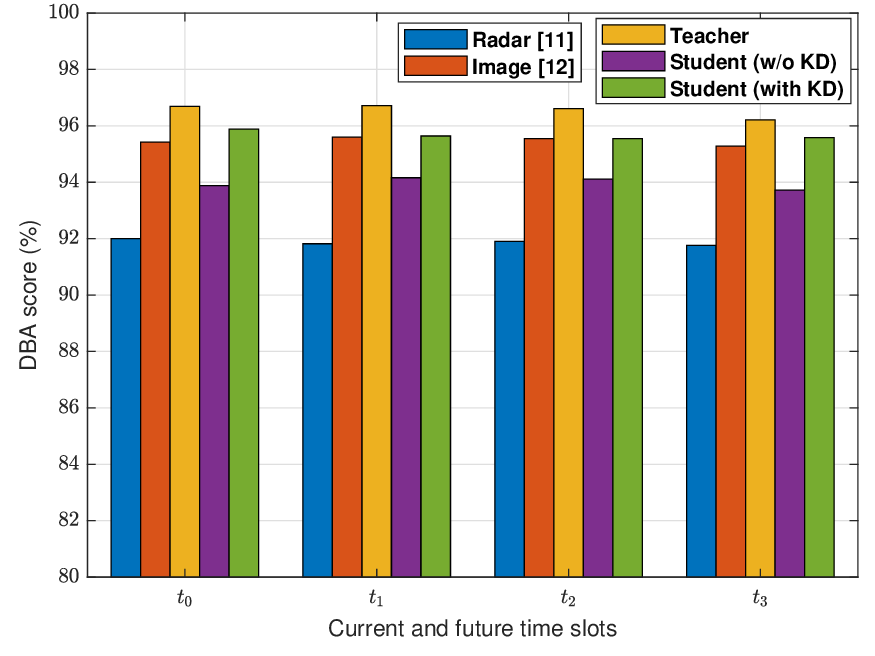}
	\vspace{-2mm}
	\caption{DBA score.}
	\label{fig:DBA}
\end{figure}

\begin{table}[t]
\centering
\caption{Model complexity comparison.}
\label{tb:model_complexity}
\begin{tabular}{l|c|c}
\hline
Model & Params (M) & FLOPs (M) \\
\hline
Radar \cite{Luo2023millimeter} & $0.275$ & $404.752$ \\
Image \cite{ma2025attention} & $1.788$ & $136.915$ \\
Teacher (Image+Radar) & $2.948$ & $179.248$ \\
Student (No KD / KD) & $\bf 0.106$ ($27\times$ fewer)& $\bf 42.723$ ($4\times$ fewer)\\
\hline
\end{tabular}
\end{table}

Table~\ref{tb:overall_generalization} summarizes the overall generalization performance averaged across time slots for different sensing modalities and models. We observe that the teacher model achieves the best performance among all approaches, with ATop-$3$, ATop-$5$, and ADBA reaching $86.96\%$, $96.84\%$, and $96.55\%$, respectively. Compared with single-modality baselines \cite{Luo2023millimeter,ma2025attention}, the teacher benefits from joint image–radar information, significantly improving prediction accuracy. The lightweight student model without KD shows noticeable performance degradation due to its reduced model capacity. However, by incorporating KD, the student model recovers most of the teacher’s performance. In particular, the student with KD improves ATop-$3$ from $80.49\%$ to $84.57\%$ and ATop-$5$ from $93.57\%$ to $95.64\%$, while achieving an ADBA of $95.66\%$, which is very close to the performance of the teacher model. These results demonstrate that KD effectively transfers knowledge from the teacher to the compact student model, enabling high prediction accuracy with significantly reduced model size.
 
Figs.~\ref{fig:Top3n5}~and~\ref{fig:DBA} illustrate the beam prediction performance across the current and future time slots in terms of Top-$k$ accuracy and DBA score, respectively. In Fig.~\ref{fig:Top3n5}, the lower and upper stacked bars represent the Top-$3$ accuracy and the additional gain from Top-$3$ to Top-$5$ accuracy. We see that the teacher model consistently achieves the best performance across all time slots, benefiting from the fusion of image and radar modalities. The student model without KD shows noticeable performance degradation due to its limited model capacity. In contrast, incorporating KD significantly improves the student's prediction accuracy, allowing it to consistently outperform both single-modality baselines. Similar trends are observed in Fig.~\ref{fig:DBA}, where the student with KD achieves DBA scores close to those of the teacher across all time slots. Moreover, the performance remains relatively stable for future prediction horizons, demonstrating the robustness of the proposed KD framework for long-term beam prediction.

Table~\ref{tb:model_complexity} compares the model complexity in terms of the number of trainable parameters and floating-point operations (FLOPs). The teacher model clearly has the largest memory footprint due to its multimodal architecture. In contrast, the student has a significantly reduced model size of only $0.106$M parameters, which is about $27\times$ fewer than the teacher. Meanwhile, the computational complexity is reduced from $179.248$M to $42.723$M FLOPs, approximately a $4\times$ reduction. The student model also requires substantially lower complexity and fewer parameters than the single-modality baselines. Despite its lightweight and low-complexity design, the student achieves performance close to that of the teacher when aided by KD, as shown in Table~\ref{tb:overall_generalization} and Figs.~\ref{fig:Top3n5} and~\ref{fig:DBA}. These results demonstrate that the proposed KD framework enables low-complexity models while maintaining strong long-term beam prediction performance.

\section{Conclusions}
This paper has proposed an efficient multimodal sensing-assisted learning framework for long-term beam prediction and tracking in mmWave communications systems. By leveraging complementary information from radar and camera sensors, a CNN–GRU-based teacher network is first developed to predict current and future beam directions. The teacher model then guides the training of a lightweight student model via KD, enabling strong predictive performance with significantly reduced complexity. Numerical results demonstrated that multimodal sensing significantly improves beam prediction performance compared with single-modality approaches. Furthermore, the distilled student model achieves performance close to that of the teacher while substantially reducing model complexity and parameter size. These results highlight the effectiveness of KD for enabling practical and efficient beam management in future sensing-assisted wireless communications systems.

\bibliographystyle{IEEEtran}
\bibliography{conf_short,jour_short,refs-my}

@string{ icc = {Proc. IEEE Int. Conf. Commun.}}

@string{ icassp = {Proc. IEEE Int. Conf. Acoust., Speech, Signal Processing}}

@string{ vtc = {Proc. IEEE Veh. Technol. Conf.}}

@string{ wcnc = {Proc. IEEE Wireless Commun. and Networking Conf.}}

@string{ eusipco = {Proc. European Sign. Proc. Conf.}}

@STRING{IEEE_J_STSP       = "{IEEE} J. Sel. Topics Signal Process."}

@STRING{IEEE_J_COM        = "{IEEE} Trans. Commun."}

@STRING{IEEE_J_WCOML      = "{IEEE} Wireless Commun. Lett."}

@STRING{IEEE_M_COM        = "{IEEE} Commun. Mag."}

@ARTICLE{Jiang2024LiDAR,
  author={Jiang, Shuaifeng and Charan, Gouranga and Alkhateeb, Ahmed},
  journal=IEEE_J_WCOML,
  title={Lidar Aided Future Beam Prediction in Real-World Millimeter Wave {V2I} Communications},
  year={2023},
  volume={12},
  number={2},
  pages={212-216},
  doi={10.1109/LWC.2022.3219409}
}

@INPROCEEDINGS{Demirhan2022Radar,
  author={Demirhan, Umut and Alkhateeb, Ahmed},
  booktitle=wcnc,
  title={Radar Aided {6G} Beam Prediction: Deep Learning Algorithms and Real-World Demonstration},
  year={2022}
}

@article{jiang2021road,
  title={The road towards {6G}: A comprehensive survey},
  author={Jiang, Wei and Han, Bin and Habibi, Mohammad Asif and Schotten, Hans Dieter},
  journal=IEEE_J_OJCOMS,
  volume={2},
  pages={334--366},
  year={2021},
  publisher={IEEE}
}

@article{imran2024environment,
  title={Environment semantic communication: Enabling distributed sensing aided networks},
  author={Imran, Shoaib and Charan, Gouranga and Alkhateeb, Ahmed},
  journal=IEEE_J_OJCOMS,
  year={2024},
  volume={5},
  number={},
  pages={7767-7786},
  publisher={IEEE},
  doi={10.1109/OJCOMS.2024.3509453}
}

@inproceedings{charan2022vision,
  title={Vision-position multi-modal beam prediction using real millimeter wave datasets},
  author={Charan, Gouranga and Osman, Tawfik and Hredzak, Andrew and Thawdar, Ngwe and Alkhateeb, Ahmed},
  booktitle=wcnc,
  year={2022}
}

@inproceedings{morais2023position,
  title={Position-aided beam prediction in the real world: How useful {GPS} locations actually are?},
  author={Morais, Jo{\~a}o and Bchboodi, Arash and Pezeshki, Hamed and Alkhateeb, Ahmed},
  booktitle=icc,
  year={2023}
}

@article{alkhateeb2023deepsense,
  title={DeepSense {6G}: A large-scale real-world multi-modal sensing and communication dataset},
  author={Alkhateeb, Ahmed and Charan, Gouranga and Osman, Tawfik and Hredzak, Andrew and Morais, Joao and Demirhan, Umut and Srinivas, Nikhil},
  journal=IEEE_M_COM,
  volume={61},
  number={9},
  pages={122--128},
  year={2023},
  publisher={IEEE}
}

@article{vaswani2017attention,
  title={Attention is all you need},
  author={Vaswani, Ashish and Shazeer, Noam and Parmar, Niki and Uszkoreit, Jakob and Jones, Llion and Gomez, Aidan N and Kaiser, {\L}ukasz and Polosukhin, Illia},
  journal={Advances in Neural Information Processing Systems},
  volume={30},
  year={2017}
}

@article{park2025resource,
  title={Resource-Efficient Beam Prediction in mmWave Communications with Multimodal Realistic Simulation Framework},
  author={Park, Yu Min and Tun, Yan Kyaw and Saad, Walid and Hong, Choong Seon},
  journal={arXiv preprint arXiv:2504.05187},
  year={2025}
}

@article{cui2024sensing,
  title={Sensing-assisted high reliable communication: A transformer-based beamforming approach},
  author={Cui, Yuanhao and Nie, Jiali and Cao, Xiaowen and Yu, Tiankuo and Zou, Jiaqi and Mu, Junsheng and Jing, Xiaojun},
  journal=IEEE_J_STSP,
  volume={18},
  number={5},
  pages={782--795},
  year={2024},
  publisher={IEEE}
}

@inproceedings{lin2017focal,
  title={Focal loss for dense object detection},
  author={Lin, Tsung-Yi and Goyal, Priya and Girshick, Ross and He, Kaiming and Doll{\'a}r, Piotr},
  booktitle={Proc. IEEE International Conf. on Computer Vision},
  pages={2980--2988},
  year={2017}
}

@INPROCEEDINGS{Luo2023millimeter,
  author={Luo, Hao and Demirhan, Umut and Alkhateeb, Ahmed},
  booktitle=eusipco,
  title={Millimeter Wave {V2V} Beam Tracking using Radar: Algorithms and Real-World Demonstration},
  year={2023}
}

@article{howard2017mobilenets,
  title={Mobilenets: Efficient convolutional neural networks for mobile vision applications},
  author={Howard, Andrew G and Zhu, Menglong and Chen, Bo and Kalenichenko, Dmitry and Wang, Weijun and Weyand, Tobias and Andreetto, Marco and Adam, Hartwig},
  journal={arXiv preprint arXiv:1704.04861},
  year={2017}
}

@inproceedings{ma2025attention,
  title={Attention-enhanced learning for sensing-assisted long-term beam tracking in {mmWave} communications},
    author={Ma, Mengyuan and Nguyen, Nhan Thanh and Shlezinger, Nir and Eldar, Yonina C and Juntti, Markku},
  booktitle=icassp,
  year={2026}
}

@article{ma2025knowledge,
  title={Knowledge Distillation for Sensing-Assisted Long-Term Beam Tracking in {mmWave} Communications},
  author={Ma, Mengyuan and Nguyen, Nhan Thanh and Shlezinger, Nir and Eldar, Yonina C and Swindlehurst, A Lee and Juntti, Markku},
  journal={arXiv preprint arXiv:2509.11419},
  year={2025}
}

@inproceedings{marasinghe2022lidar,
  title={Lidar aided wireless networks-beam prediction for {5G}},
  author={Marasinghe, Dileepa and Jayaweera, Nalin and Rajatheva, Nandana and Hakola, Sami and Koskela, Timo and Tervo, Oskari and Karjalainen, Juha and Tiirola, Esa and Hulkkonen, Jari},
  booktitle=vtc,
  year={2022}
}

@article{zhu2025advancing,
  title={Advancing multi-modal beam prediction with cross-modal feature enhancement and dynamic fusion mechanism},
  author={Zhu, Qihao and Wang, Yu and Li, Wenmei and Huang, Hao and Gui, Guan},
  journal=IEEE_J_COM,
  year={2025},
  volume={73},
  number={9},
  pages={7931-7940},
  publisher={IEEE}
}

\end{document}